# Nanoindentation stress-strain for Fracture Analysis and computational modeling for hardness and modulus

A.S.Bhattacharyya, S, Priyadarshi, Sonu, S. Shivam, S. Anshu

Centre for Nanotechnology
Central University of Jharkhand
Brambe, Ranchi: 8352005

arnab.bhattacharya@gmail.com

Abstract

Stress-Strain plots based on Nanoindentation load-depth curves were plotted. Cracking phenomena during the indentation process were analyzed based on the stress–strain plots. A transition from ductile to brittle fracture was observed on increasing the depth or load of indenter penetration. A new approach with shape factors in the fracture studies based on radial crack branching and micro-cracking was done. Hardness and modulus plots were fitted with polynomials. The coefficients were varied to obtain different hardness and modulus responses.

Keywords: Nanoindentation, load-depth, stress-strain, fracture

Nanoindentation has been a promising tool to determine hardness and other mechanical properties at nanoscale. It is based upon depth sensing and continuous stiffness mode. Both bulk modulus (E) and hardness (H) are found from nanoindentation [1, 2]. The advantage of this technique lies in the possibility of very small indentation (depth of the order of 100 nm) [3]. Thus it is useful in the case of thin films. The indentations were carried out by Nanoindenter XP (MTS, USA) on Ti-B-Si-C hard coatings developed on Si substrates.

During nanoindentation a Berkovich indenter with 70.3° effective cone angle pushed into the material and withdrawn. The shape of the indentation is triangular as shown in Fig 1. The bright area surrounding the indentation is due to pile-up during plastic deformation.

The hardness and elastic modulus are calculated simultaneously using a method developed by Oliver and Pharr [4]. The advantage of using a sharp 3 sided pyramidal indenter (Berkovich) is the plastic deformation starts in the coating at very low loads and the size of the plastic zone increases as the load increases. Nanoindentation is useful for a coating/substrate system as we can obtain the mechanical property of only the coating material eliminating the substrate effect.

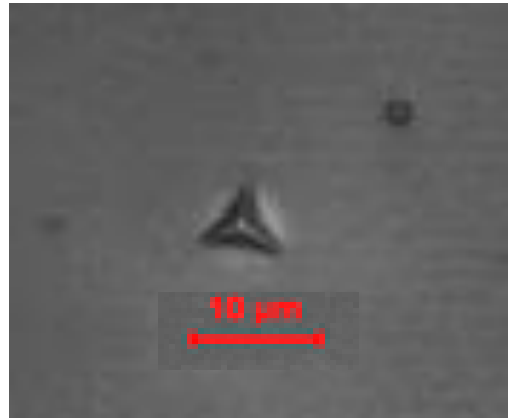

Fig 1: Nanoindentation by Berkovich indenter

For engineering applications, hardness must be complimented with high toughness, which is a property of equal importance as hardness. Toughness is an important mechanical property related to the materials resistance against the formation of cracks. In an energetic context, toughness is the ability of a material to absorb energy during deformation up to fracture. Nanoindentation performed at higher loads caused fracture surrounding the indentation impression [5-7]. However there are also some internal cracks not visible on the surfaces which arise due to high shear stress of the indenter. Evidence of these internal cracks can be found in the discontinuity of the load depth curve or corresponding stress-strain plots [8] as shown in Fig 2.

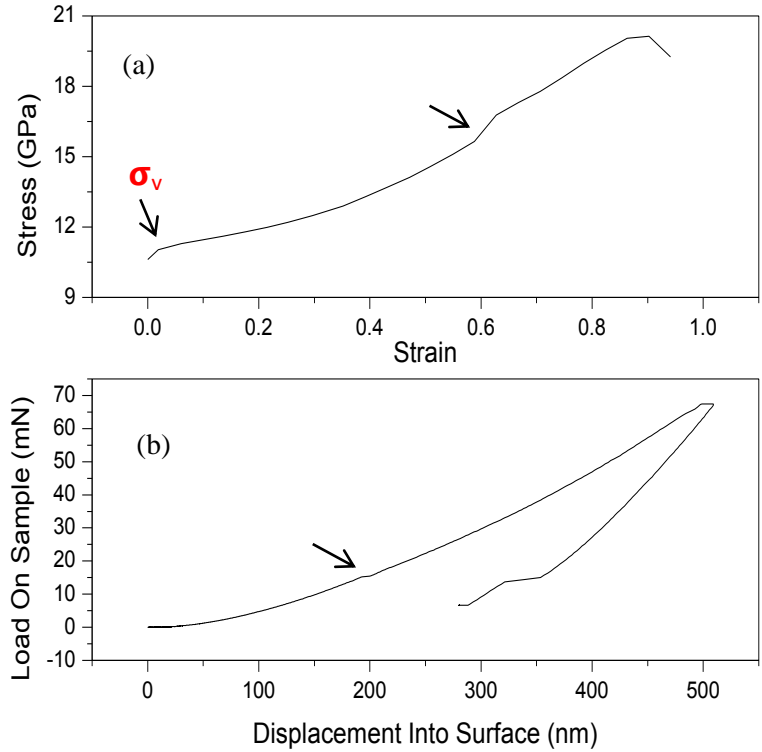

Fig 2: Stress-Strain and load depth curves for a penetration of 500nm

Stress-strain plots are also plotted for a higher penetration depth for the same material (Fig 3). Interestingly a lower stress was found to cause the same amount of strain for the second case. The reason is effect of comparatively softer substrate. The yield points ($\sigma_y$) are indicated in the figure.

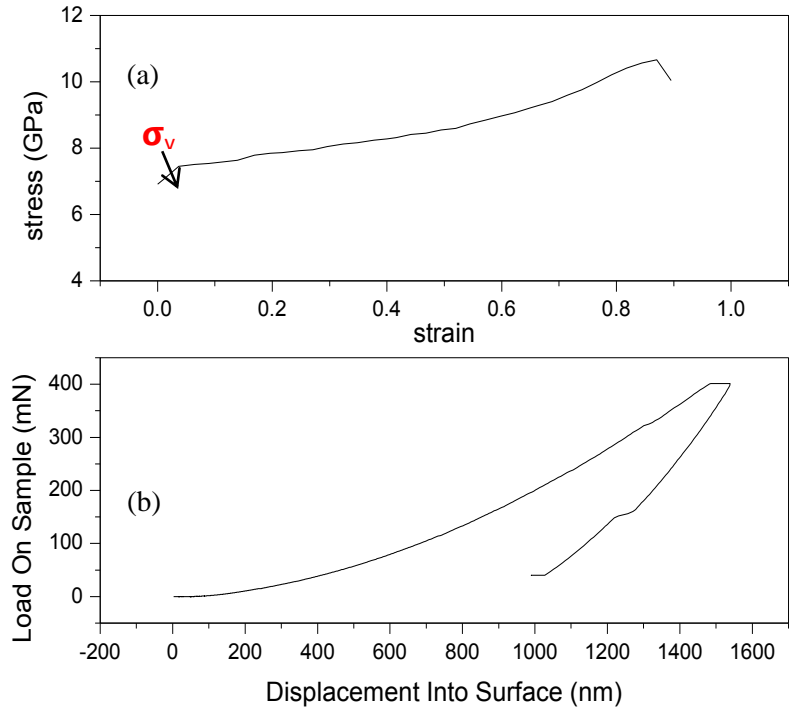

Fig 3: Stress-Strain and load depth curve for a penetration of 1500nm

The discontinuity due to fracture is sometimes also not visible in the stress-strain plot as observed for indentations done at higher penetration. The discontinuity becomes more prominent if we take the derivative of the stress strain curve as shown in Fig 4 for both the cases. It can be seen that for 500 nm penetration depth a prominent cracking takes place in the strain range of 0.6 – 0.7 On the other hand the internal cracking phenomenon was different for 1500 nm with no major but multiple cracking at strains of 0.18, 0.3, 0.5 and 0.8.

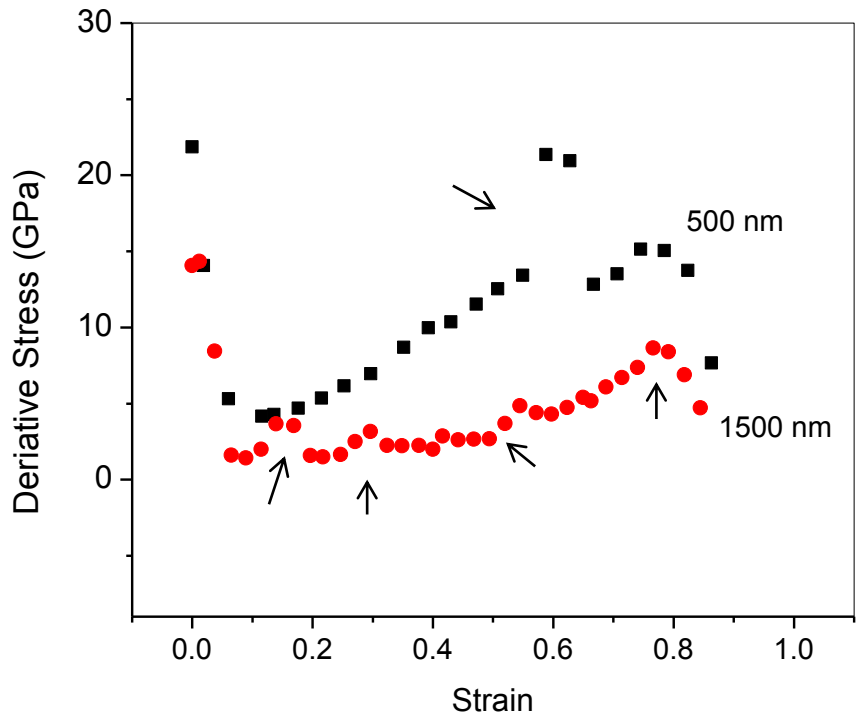

Fig 4: Derivative of Stress-Strain plot at 500nm and 1500 nm

On looking at the nanoindents corresponding to the above mentioned two indentations we can see that no prominent cracks on the surface is visible for 500nm indentation unlike indentations at 1500 nm where radial cracks along with lateral cracks and chipping is observed.

As both plastic deformation and fracture are involved in nanoindentation, it can be considered similar to ductile fracture where appreciable plastic deformation occurs before crack initiation and during crack propagation. The Berkovich indenter is specially made to cause plastic deformation even at shallow depths. The prominent discontinuity for 500 nm depth penetration can be an example of ductile fracture where the cracks are internal and have not been able to propagate to the surface.

However as there is no clear demarcation between ductile and brittle fracture during indentation, the percentage of ductile fracture reduces and brittle fracture increases with high load or depth of penetration. The multiple discontinuity are then due to this brittle fracture where the cracks propagate to the surface and come out as radial cracks from the indentation corners and also undergo chipping as observed in Fig 5.

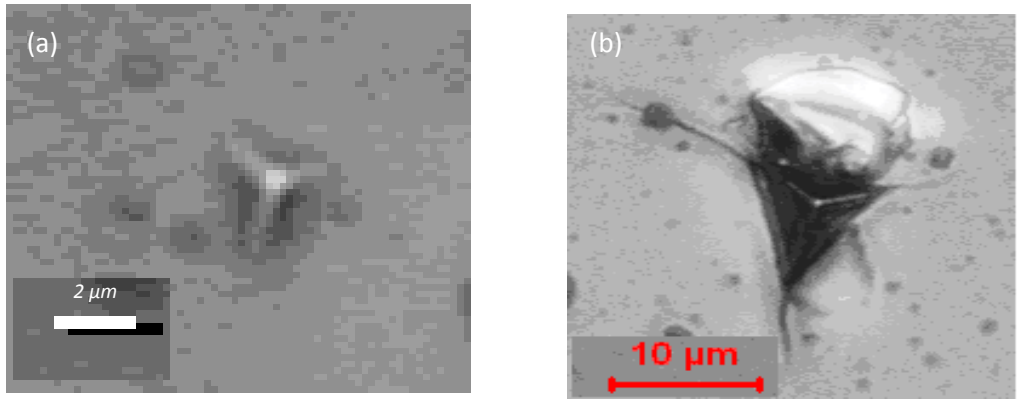

Fig 5: Nanoindentation at a) 500nm and b) 1500 nm depth

Presence of uniaxial stress usually causes the material to deform plastically without fracturing due to large shear stress. Brittle fracture however occurring is associated with tri-axial state of stress [9]. Hence during the indentation process a tri-axial stress acts although the applied stress is unidirectional. The tendency of brittle fracture increases with increasing strain rate. Although a fixed indentation strain rate of $0.05s^{-1}$ is used during the indentation, it is the shear strain which increases with increase in depth of penetration causing brittle fracture.

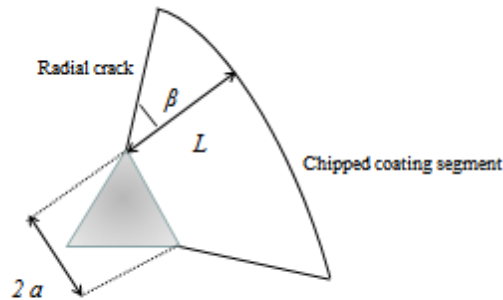

Fig 6: Geometrical representation of chipped region surrounding indentation [10]

The failure region after radial cracks gets converted into a chipped region from any one side of the indentation impression. A per Bull, the geometrical parameters in the chipped segment as shown in Fig 6 can be used to determine the interfacial fracture energy [10]. The values of a, **β** and L come out to be roughly 3µm, 30º and 6µm. Using these values in the equation below as proposed by den Toonder et al [11], we obtained the interfacial fracture energy $\Gamma_i$ more than 10 J m$^{-2}$  E (= 150- 200 GPa) is the elastic modulus, $t$ (1 - 3 µm) the coating thickness, **σ** the residual stress (50 – 100 MPa), ν the Poisson's ratio (0.25).

$$\Gamma_i = 1.42 \ \frac{Et^5}{L^4} \left(\frac{\frac{\alpha}{L}+\frac{\beta\pi}{2}}{\frac{\alpha}{L}+\beta\pi}\right)^2 + \frac{t(1-\nu)\sigma^2}{E} + \frac{3.36((1-\nu)t^3\sigma}{L^2} \left(\frac{\frac{\alpha}{L}+\frac{\beta\pi}{2}}{\frac{\alpha}{L}+\beta\pi}\right)$$

The impulse plays a major role in indentation fracture. Indentation with the same load but different impulse (due to different time on sample) may cause different failure responses [12]. A lower impulse will lead to radial crack formation whereas higher impulse indentation will provide higher shock wave which gets reflected from the films surface interface making chances of buckling, delamination and chipping more. The effect of substrate will be also much higher for indentations done at higher impulse. This is the reason while we get radial cracks during loading when the impulse is low and lateral crack during unloading when the impulse is high [12, 13].

There exists a characteristic inelastic volume just beneath the sharp indenter where compressive stresses provide resistance to crack propagation. An increase in load however makes the cracks grow faster than inelastic volume. Although radial cracks may get deviated due to crystallographic orientation, the lateral cracks follow the stress field due to shock waves and are not affected by crystallography. A crystallographic anisotropy also leads to extra crack generation and the energy is not spent in crack propagation [14]. These high impulse shockwaves are the reason of flow of material surrounding the indentation impression which takes different geometrical shapes. However due to improper sampling, variation in thickness, the shockwaves may not propagate with equal intensity back to the surface on being reflected which makes the failure region surrounding the indenter impression inhomogenous as shown in Fig 7 with three geometrical fractured zone indicated as 1, 2 and 3 surrounding the indentation due buckling. However the height of the fractured zones from the surface was different giving them different brightness. Lateral cracks were the boundary of the zones. It was discussed previously that the fractured zones or chipped coating segments can be analyzed to determine the interfacial toughness. However, the interfacial toughness may not be isotropic in nature as the morphology of the chipped regions are not uniform.

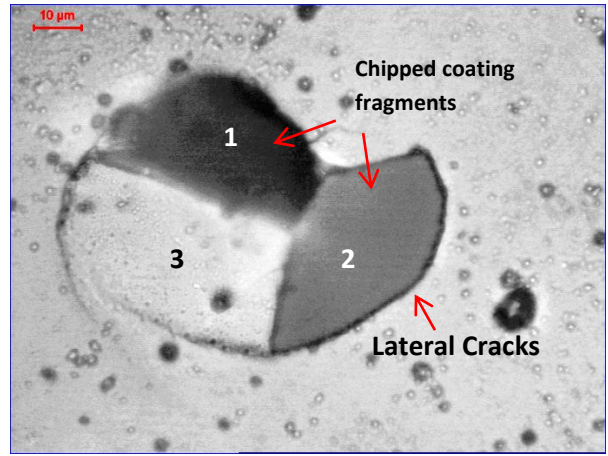

**Fig 7**: Lateral crack and buckling into three zones during indentation

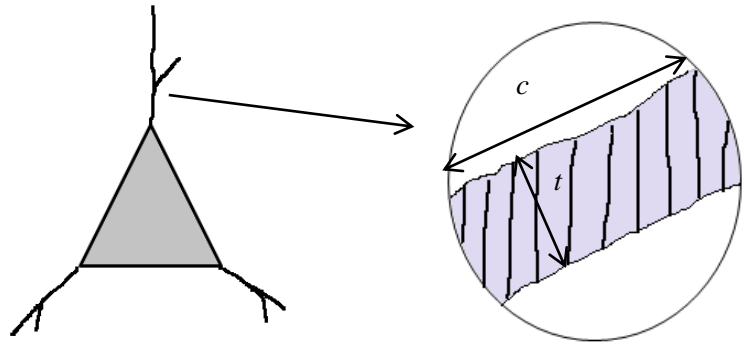

**Fig 8:** Crack branching / Microcracks during Nanoindentation [12]

The radial cracks obtained during Vicker's Indentation were initially used to determine fracture toughness. The phenomenon of stress induced crystallization has been found to cause branches or microcracks in the radial cracks (Fig 8). According to Moradkhani et.

al the micro cracks can be assumed to have a regular shaped geometry and used in fracture toughness calculation. The area has been taken as rectangular with length c and thickness t such that c=A/t where A is the micro crack area (Fig 9). The fracture toughness is then given by

$$K_{IC} = \chi \frac{P}{c^{3/2}}$$

Where the constant $\chi = \zeta (E/Hv)^{1/2}$ where $\zeta$ is a dimensionless empirical constant having value 0.016 approximately [12, 15]. A study relating the fracture toughness with indentation time has been previously reported [12]. To extend the research further we assigned other geometrical shapes to the crack region. We observed crack branching phenomenon in nanoindentation as well as reported earlier [16]. The ratio E/H is nanoindentation is usually of the order of 10 which gives $\chi \sim 0.05$. The initiation point of branching can be assigned a triangular or circular shape as shown below (Fig 10)

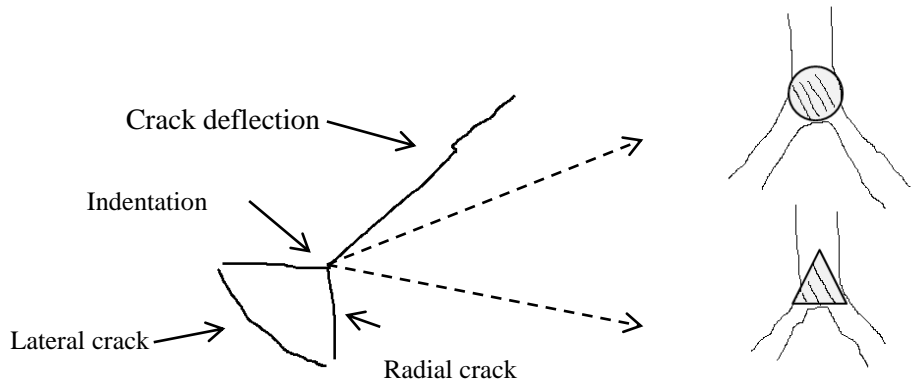

**Fig 9:** Crack branching / Microcracks taking circular and triangular shape during Nanoindentation [16]

The crack length for the spherical case can be taken as the diameter of the circle whereas for the triangle can be taken as one of the sides. The variation of fracture toughness variation with respect to applied load and crack length were estimated computationally and given in Fig 12.

Similar to nanoindentation formation of lateral and radial cracks also take place during microindentation (Fig 10)

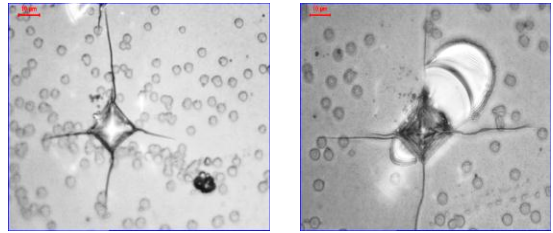

**Fig 10:** Radial and Lateral cracks during Vickers micro indentation

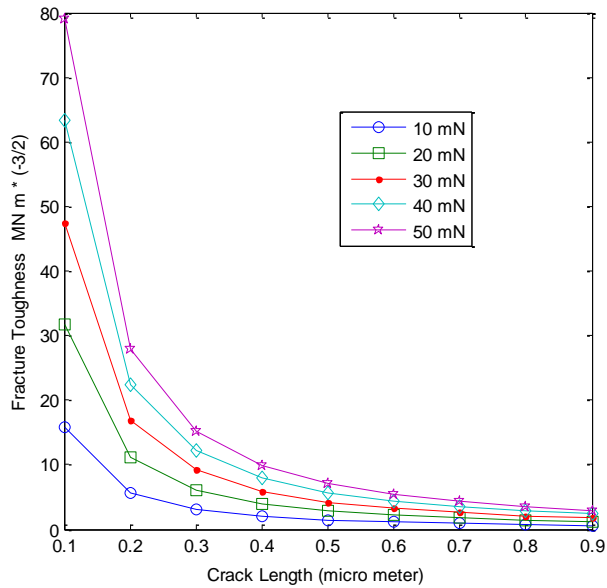

**Fig 11**: Fracture toughness variation with crack length and load

However, if we consider spherical or triangular shapes, a shape factor should also be introduced in the fracture toughness calculation. We consider the shape factor as the area ratio w.r.t a square with side length being equal to diameter length in case of circle and side length in case of triangle, which gives a shape factor of $\pi/4$ in case of circle and $\sqrt{3}/4$ in case of triangle considering it to be an equilateral one. The fracture toughness obtained for the three cases is shown comparatively in Fig 12 for a load of 10 mN.

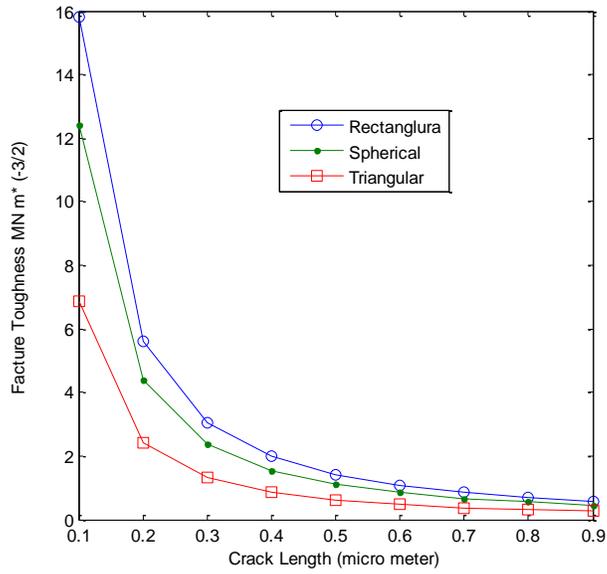

**Fig 12.** Fracture toughness variation with crack length and load assuming different geometrical shapes of the micro crack region.

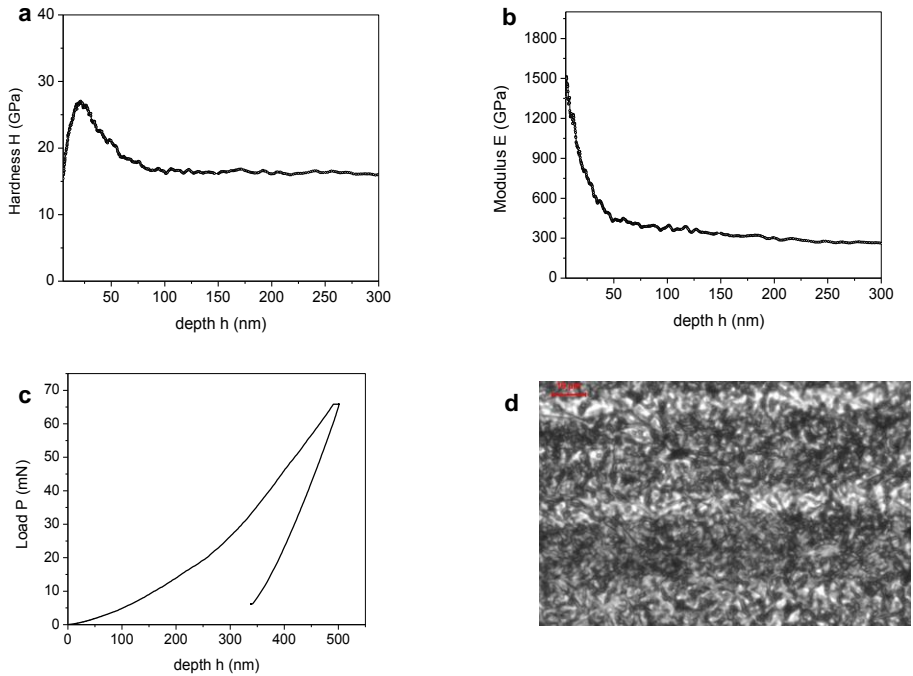

Fig 13 : Nanoindentation of TiN (a) Hardness (b) Modulus (c) load-depth (d) surface structure

Nanoindentation performed on nitrogen Plasma surface modified Titanium is shown in Fig 13 where a high hardness of 27 GPa was obtained due to formation of TiN. Nanoindentation of Si-C-N coatings deposited on Stainless Steel (SS304) showed a hardness of about 14 GPa and modulus of 160 GPa modulus as shown in Fig 14 and 15 respectively. The lowering of values after 250 nm is due to substrate effect which starts at 10% of the coating thickness making the coating 2.5µm thick. The load-depth curve showed a high plastic area due to underlying SS304 substrates (Fig 16). The hardness and modulus plots with depth of penetration were computationally fitted with a mathematical relation as given in eqn 1 and 2 with coefficients given in Table 1.

```
H = p1*h^7 + p2*h^6 + p3*h^5 + p4*h^4 + p5*h^3 + p6*h^2 + p7*h + p8
............ (1)

E = q1*h^10 + q2*h^9 +  q3*h^8 + q4*h^7 + q5*h^6 + q6*h^5 + q7*h^4 +
q8*h^3 + q9*h^2 + q10*h +    q11 ......... (2)
```

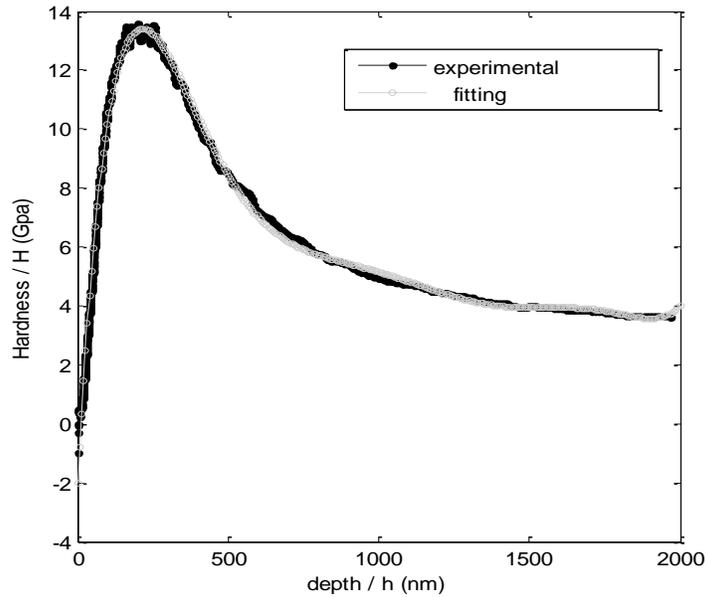

Fig 14: Experimental and computational fitted Hardness profile

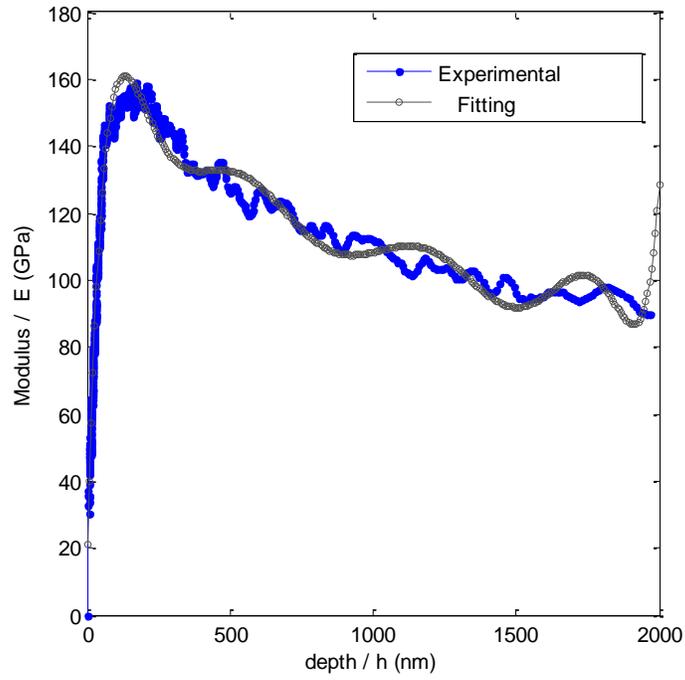

Fig 15: Experimental and computational fitted modulus profile

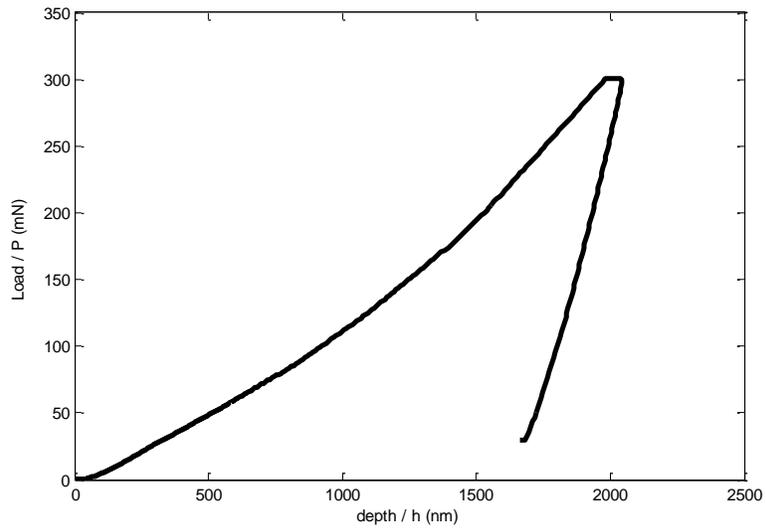

Fig 16: Load-depth curve for nanoindntation of SiCN on SS304 substrates

**Table 1: Coefficients related to fitting of hardness and modulus plots**

| p1= | 3.5374e-20 | q1 | -3.7261e-27 |
|---|---|---|---|
| p2= | -2.8657e-16 | q2 | 3.7912e-23 |
| p3= | 9.4833e-13 | q3 | -1.6563e-19 |
| p4= | -1.6436e-09 | q4 | 4.0626e-16 |
| p5= | 1.5833e-06 | q5 | -6.1417e-13 |
| p6= | -0.00081871 | q6 | 5.9165e-10 |
| p7= | 0.18908 | q7 | -3.6306e-07 |
| p8= | -2.0343 | q8 | 0.00013795 |
|  |  | q9 | -0.030446 |
|  |  | q10 | 3.3698 |
|  |  | q11 | 16.272 |

The coefficients were varied using MATLAB codes to get different hardness and modulus responses. We observed that variation in p8 and q11 were affecting the peak hardness and modulus values as shown in Fig 17 and 18 respectively As hardness cannot be less than zero so a reference line in drawn at H= 0 GPa below which the values are not considered.

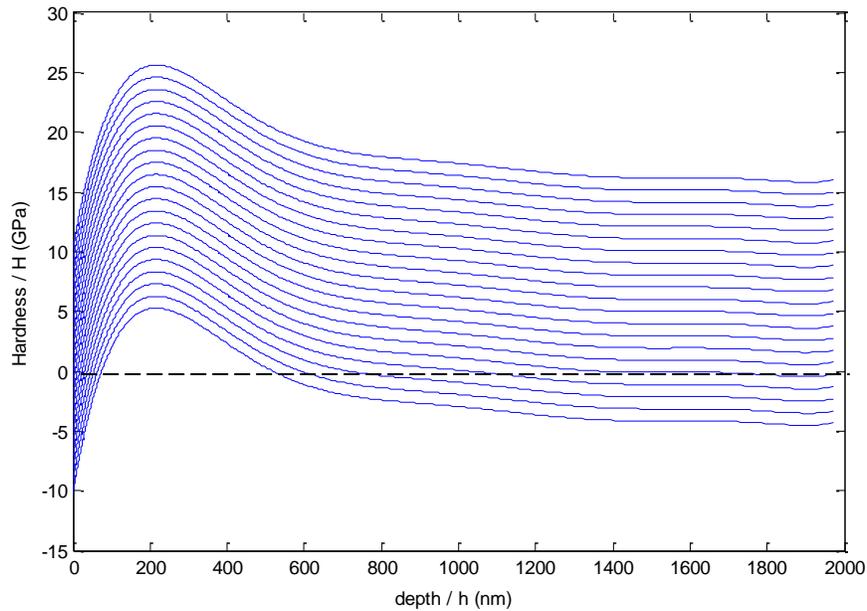

Fig 17: Hardness profile with variation in p8

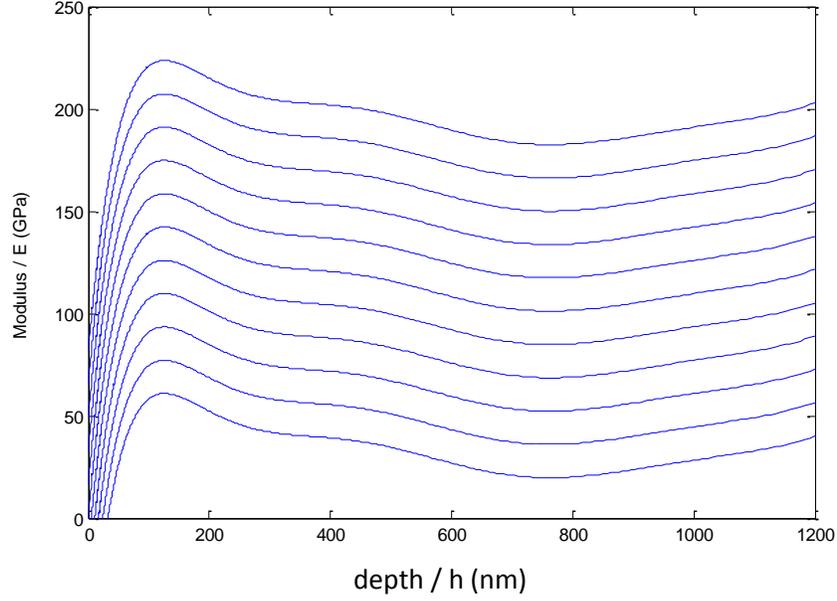

Fig 18: Modulus profile with variation in p8

Hardness and Modulus usually depend on the cohesive strength. As H and E vary with p8 and q11. Hence the coefficients can be related to the cohesive strength. High cohesive forces are related to large elastic constants, high melting point and small coefficients of thermal expansions. The expression of cohesive strength is given as $E/\pi$ for brittle elastic solid and changes to $\sigma_c = \left(\frac{E\gamma_s}{a_o}\right)^{1/2}$ when fracture occurs where E is the modulus, $\gamma_s$ is the surface energy and $a_o$ is the interatomic spacing in the unstrained condition [9].

The next step was to other coefficients along with p8 for each case. We obtained variation in peak hardness value although no variation in the shape of the curve was observed (Fig 19). Variation of powers of h along with the coefficients however caused some changes in the position of the peak value as shown by the arrow (Fig 20) indicating an increase in film thickness.

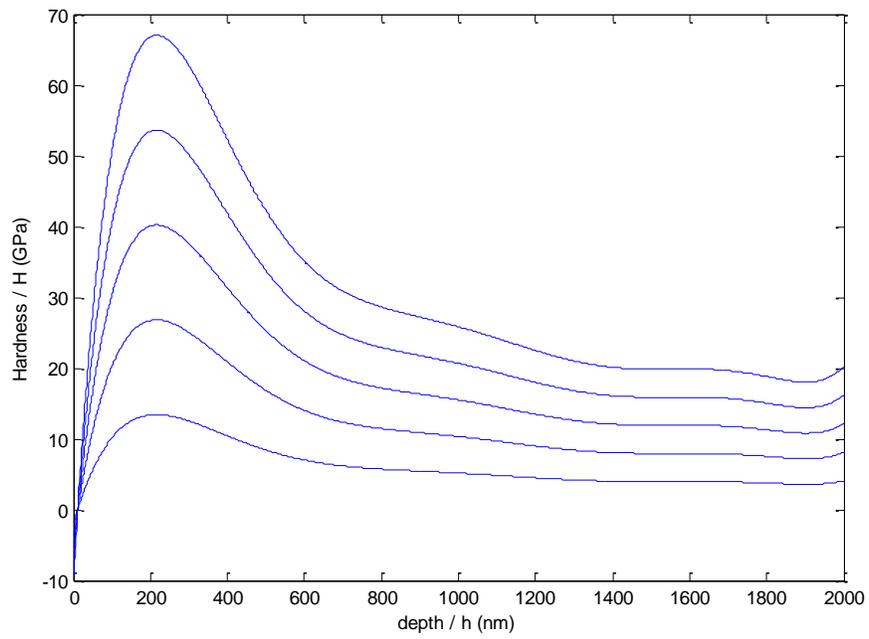

Fig 19: Variation of H with varying all the coefficents

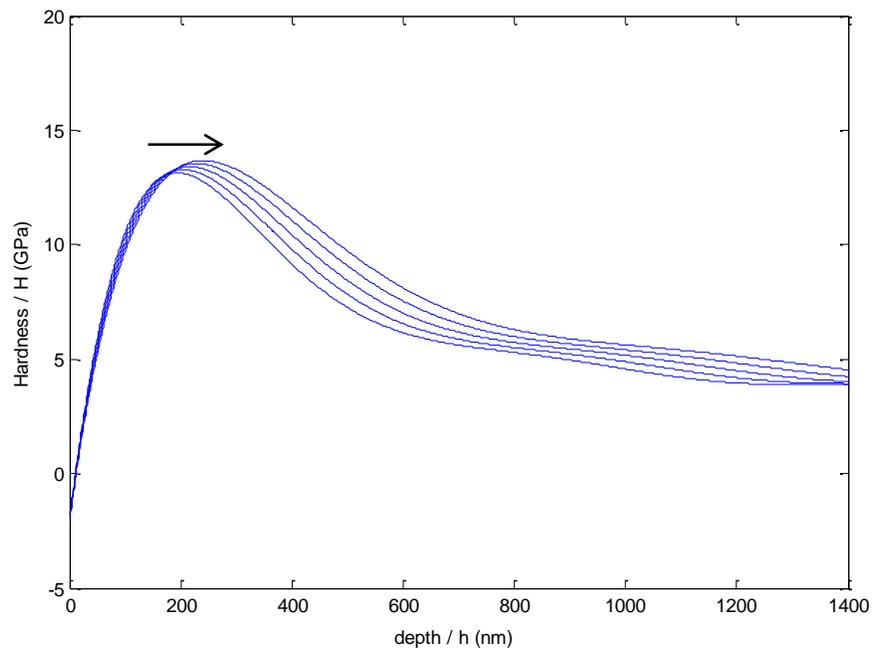

Fig 20: Variation of H with varying all the coefficients and powers of h

## Conclusions

Stress-Strain plots were drawn for nanoindentation load depth curves to analyze the internal cracking phenomenon during indentation. A higher loading resulted in multiple cracks at lower strains. Ductile to brittle fracture were observed with increasing the penetration depth. Fracture toughness of the coatings was also studied based upon the cracks developed and failure modes during the mechanical tests. Nanoindentation was performed on hard surfaces like nitrogen plasma modified Ti and Si-C-N. The hardness and modulus plots were computationally fitted with mathematical equations. The coefficients of the fitted polynomial was varied to get films of different hardness and thickness. The results will provide a mathematical frame work to the indentation tests.